\documentclass[12pt]{article}
\usepackage{amsfonts}
\usepackage{amssymb,amsmath,amsthm,latexsym}
\usepackage[numbers,sort&compress]{natbib}
\usepackage{amsmath}
\usepackage{indentfirst}
\allowdisplaybreaks[4]
\newtheorem{theorem}{Theorem}[section]
\newtheorem{prop}[theorem]{Proposition}
\newtheorem{cor}[theorem]{Corollary}
\newtheorem{lemma}[theorem]{Lemma}

\newtheorem{remark}[theorem]{Remark}
\newtheorem{define}[theorem]{Definition}
\newtheorem{example}[theorem]{Example}

\newcommand{\pf} { {\rm \noindent{\bf Proof.}} }

\numberwithin{equation}{section}

\begin{document}

\title{Entanglement-assisted quantum codes from Galois LCD codes }
\author{
 Xiusheng Liu {\thanks{Corresponding author. \newline  Email addresses: lxs6682@163.com(Xiusheng Liu), hwlulu@aliyun.com(Hualu Liu), longyuhbpu@163.com(Long Yu)} }, Hualu Liu, Long Yu}
\date{School of Mathematics and Physics, Hubei Polytechnic University, Huangshi 435003, China }

\maketitle


\begin{abstract}
Entanglement-assisted quantum error-correcting codes (EAQECCs) make use of preexisting entanglement
between the sender and receiver to boost the rate of transmission. It is possible to construct an EAQECC from
any classical linear code, unlike standard quantum error-correcting codes, which can only be constructed from dual-containing codes. However, the parameter of ebits
$c$ is usually calculated by computer search. In this paper, we construct four classes of MDS entanglement-assisted quantum error-correcting  codes (MDS EAQECCs)
based on $k$-Galois  LCD  MDS codes for some certain code lengths, where the parameter of ebits $c$ can be easily generated algebraically and not by computational search. Moreover, the constructed four classes of EAQECCs  are also maximal-entanglement EAQECCs.
\end{abstract}


\bf Key Words\rm :  Entanglement-assisted quantum codes, $k$-Galois LCD codes,  Parity-check matrix

\section{Introduction}
Quantum error-correcting codes play an important role in  quantum communications and quantum computations. After the pioneering  work in \cite{AK01}, \cite{Calder1},  the theory of quantum codes has developed rapidly in recent decade years. As we know, the approach of constructing new quantum codes which have good parameters is an interesting research field.   Many good  quantum codes have  been constructed by classical linear codes with Hermitian dual containing, which  can be found in \cite{AKS07}, \cite{CLZ15}, \cite{JLLX10}, \cite{JX14}, \cite{KZ13}, \cite{KZ14}, \cite{G11}, \cite{Liu1}.

An important discovery in the area of quantum error correction was the development of a
theory of entanglement-assisted quantum error-correcting  codes(EAQECC, for short). In this theory it is assumed that in addition to a quantum channel, the sender and receiver share a ceratin amount of pre-existing entanglement. EAQECCs allow the use of arbitrary classical codes (not necessarily Hermitian dual containing) for quantum data transmission via pre-shared entanglement bits (ebits). Fujiwara et al. \cite{Fuji} gave a general method for constructing entanglement-assisted quantum low-density parity check (LDPC) codes.  Fan, Chen and Xu \cite{FanJ} provided a construction of entanglement-assisted quantum maximum distance separable (MDS) codes with a small number of pre-shared maximally entangled states.

Linear complementary dual codes (abbreviated to LCD codes)
are linear codes that meet their duals trivially.
Massey \cite{Massey} introduced LCD codes,
showed the asymptotically good property of LCD codes,
and provided an optimum linear coding solution for the two-user binary adder channel. In \cite{Liu2}, we introduced the so-called $k$-Galois LCD codes which include the usual LCD codes and Hermitian LCD codes as two special cases, and gave sufficient and necessary conditions for a code to be a $k$-Galois LCD code.
Recently, Qian,  Zhang \cite{Qian}, and Guenda, Jitman, Gulliver \cite{Guend} exhibited an application of LCD codes
in constructing good EAQECCs.

Inspired by these works, in this paper, we construct several classes of  MDS EAQCCs
based on $k$-Galois  LCD MDS codes,  and  get new MDS EAQCCs with maximal-entanglement. Throughout this paper, $\mathbb{F}_{q}$ denotes
the finite field with cardinality $|\mathbb{F}_{q}|=q=p^e$,
where $p$  is a prime and $e$ is a positive integer.  An EAQECC over $\mathbb{F}_{q}$ with parameters $[[n,l,d;c]]_q$, encodes $l$ logical qubits into $n$
physical qubits using $c$ copies of maximally entangled Bell states, and $d$ is the minimum
distance of the code. The code can correct up to at least $\lfloor(d- 1)/2\rfloor$ errors acting on the $n$
channel qubits.  Our main results are to obtain  four classes of  with parameters:

$(i)$ $[[n,l,n-l+1;n-l]]_q$ or $[[n,n-l,l+1;l]]_q$, where $\gcd(p^{e-k}+1,p^e-1)=s$, $p^e-1=st$, $ n\leq q+1$, and $0\leq l<\min\{t,n\} $.

$(ii)$  $[[n,n+1-d,d;d-1]]_q$ , where $q=p^e$  and $p$ is an odd  prime,  $n=\frac{q-1}{r}$, $r|(q-1)$,  $r\nmid(1+p^{e-k})$,  and $2\leq d \leq n$.

$(iii)$ $[[2l,l,l+1;l]]_q$, $[[2l+1,l,l+2;l+1]]_q$, or  $[[2l+2,l,l+3;l+2]]_q$, where $q=p^e$ and $p$ is an odd  prime, $l$ be a positive integer with $l\mid(p^e- 1), l\nmid p^k + 1, 2(p^k+1)\mid(p^e- 1)$, and $l\leq\frac{p^e-1}{2}$.

$(iv)$ $[[(p-1)/2,(p-1)/2-2l,2l+1;2l]]_q$, where $q=p^e, e\geq2$, $p\equiv1\pmod{4}$ and $1\leq l\leq(p-5)/4$ or $[[(p-1)/2,(p-1)/2-2l+1,2l;2l-1]]_q$, where $q=p^e, e\geq2$, $p\equiv3\pmod{4}$ and $1\leq l\leq(p-3)/4$.

The following three classes of MDS EAQCCs can be found in \cite{FanJ}.

$(1)$ $[[p^{2a} + 1,p^{2a}-2d + 4,d;1]]_q$, where $q=p^{2a}$  and $p$ is a prime, $2 \leq d \leq 2p^a$ is an even integer.



$(2)$ $[[\frac{p^{2a}-1}{2},\frac{p^{2a}-1}{2}-2d + 4,d;2]]_q$ , where $q=p^{2a}$ and $p$ is an odd prime, $\frac{p^a+1}{2}+1\leq d \leq \frac{3}{2}p^a-\frac{1}{2}$.

$(3)$ $[[\frac{p^{2a}-1}{r},\frac{p^{2a}-1}{r}-2d + r+2,d;r]]_q$ ,   where $q=p^{2a}$ and $p$ is an odd prime with $r\mid(p^a + 1), r\geq 3$ is
an odd integer, and $\frac{(r-1)( p^a+1)}{2}+2\leq d \leq \frac{(r+1)( p^a+1)}{2}-2$.

Taking $e=2a$ and $n=q+1=p^{2a}+1$, the minimum distance of the above type (i), for $1\leq l \leq p^{2a}-2p^a+2$,   is greater than $2p^a$. Comparing with the type (1), our results are better.

Note that the type (ii), it is easy to see that even $r$ exists. This implies that the type (ii)  is different from the  type (2) and (3)

We finish this introduction with a description of each section in this paper. Section $2$ recalls the basics about linear codes, $k$-Galois LCD codes and EAQECCs. In Section $3$,  we give  new constructions of  EAQECCs by using $k$-Galois LCD codes.  Finally, a brief summary of this work is described in section 4.

\section{Preliminaries}
Starting from this section till the end of this paper,
we assume that $n$ is a positive integer coprime to $q$,
and by $\mathbb{F}_{q}^{*}$ we denote the multiplicative group
of units of $\mathbb{F}_{q}$.
Let $\mathbb{F}_{q}^n=\{{\bf x}=(x_1,\cdots,x_n)\,|\,x_j\in \mathbb{F}_{q}\}$
which is an $n$ dimensional vector  space over $\mathbb{F}_{q}$.
Any subspace $C$ of $\mathbb{F}_{q}^n$ is called a linear code
of length $n$ over $\mathbb{F}_{q}$.
We assume that all codes in this paper are linear.
A linear code $C$ of length $n$ over $\mathbb{F}_{q}$ is called a maximum distance separable
(abbreviated to MDS) code if its parameters $[n,l,d]_q$ attain
the singleton bound $d= n-l+1$, where $l=\dim C$ and $d$ denotes
the minimum Hamming distance of $C$.

We assume that
$\lambda\in\mathbb{F}_{q}^{*}$ with
$\mathrm{ord}_{\mathbb{F}_{q}^{*}}(\lambda)=r$,
where $\mathrm{ord}_{\mathbb{F}_{q}^{*}}(\lambda)$ denotes
the order of $\lambda$ in the  group $\mathbb{F}_{q}^{*}$, hence $r\mid q-1$.

A linear code $C$ of length $n$ over $\mathbb{F}_{q}$ is said to
be $\lambda$-constacyclic if $(\lambda c_n ,c_1 ,\ldots,c_{n-1} ) \in C$ for every $(c_1 ,c_2 ,\ldots,c_n ) \in C$. If $\lambda = 1$, $C$ is a cyclic code. If $\lambda =-1$, $C$ is called a negacyclic code.  A codeword $(c_1 ,c_2 ,\ldots,c_n ) \in C$ is identified with its polynomial representation $c(x) = c _1 +
c _2 x + ¡¤~\cdots + c_nx^n$. It is easy to find that a $\lambda$-constacyclic code $C$ of length $n$ over $\mathbb{F}_{q}$ is
an ideal of the quotient ring $\mathbb{F}_{q} [x]/\langle x^n-\lambda\rangle$. It is known that $C$ is generated by a monic
divisor $g(x)$ of $x^n-\lambda$. The polynomial $g(x)$ is called the generator polynomial of the code
$C$.

By $\mathbb{Z}_{rn}$ we denote the residue ring of
the integer ring $\mathbb{Z}$ modulo $rn$.
Let $1+r\mathbb{Z}_{rn}$ denote the following subset of $\mathbb{Z}_{rn}$:
$$1+r\mathbb{Z}_{rn}=\{1+ri\!\!\pmod{r n}\mid i\in\mathbb{Z}_{rn}\}
  =\{1,1+r,\ldots,r(n-1)\}.$$

Let $m$ be the  multiplicative order of $q$ modulo $rn$,
i.e., $rn\mid(q^{m}-1)$ but $rn\nmid(q^{m-1}-1)$.
Then in $\mathbb{F}_{q^{m}}$ there exists a primitive
$rn$-th root $\theta$ of unity such that $\theta^{n}=\lambda$.
It is easy to check that $\theta^{i}$, $i \in 1+r\mathbb{Z}_{rn}$,
are all roots of $x^n-\lambda$.
In $\mathbb{F}_{q^m}[x]$,  we have:
$$x^n-\lambda=\prod_{i\in(1+r\mathbb{Z}_{rn})}(x-\theta^{i}).$$

The defining set of the  $\lambda$-constacyclic code $C$  with generator
polynomial $g(x)$ is defined as
$$P=\{1+ir\in 1+r\mathbb{Z}_{rn}\mid g(\theta^{1+ir})=0 \}.$$
Obviously, the defining set $P$ is a union of
some $q$-cyclotomic cosets modulo $rn$ and
$$\mathrm{dim}(C)=n-|P|,$$
where the $q$-cyclotomic coset modulo $rn$ containing $i$ is denoted by $C_i = \{i,iq ,iq^2 ,...,iq^{m_i-1}\}$, where
$m_i$ is the smallest positive integer such that $iq^{m_i}\equiv i~(\mathrm{mod}~rn)$.

The following result is important in constructing optimal constacyclic codes (see\cite{Kri}).

\begin{prop} \label{pro:2.1}
$\mathrm{(The~BCH ~bound ~for~ constacyclic ~codes)}$ Suppose that $\mathrm{gcd}(q,n)=1$.
If the defining set of a $\lambda$-constacyclic code $C$ of length $n$ over
$\mathbb{F}_{q}$ contains a subset
$ \{1+ri\mid i=h,h+1,\cdots, h+r(\delta-2)\}$,
then the minimum distance of $C$ is at least~$\delta$.
\end{prop}

\subsection{$k$-Galois LCD codes}
In \cite{FanY}, Fan and Zhang introduced a kind of forms on $\mathbb{F}_{q}^n$
as follows: for each integer $k$ with $0\leq k< e $ , define:
$$
[{\bf x},{\bf y}]_{k}=x_1y_1^{p^{k}}+\cdots+x_ny_n^{p^{k}},
\qquad\forall~ {\bf x},{\bf y}\in\mathbb{F}_{q}^n.
$$
We call $[{\bf x},{\bf y}]_{k}$ the $k$-Galois form on $\mathbb{F}_{q}^n$.
It is just the usual Euclidean inner product if $k=0$.
And, it is  the Hermitian inner product if $e$ is even and $k=\frac{e}{2}$.
For any code $C$ of $\mathbb{F}_{q}^n$, the following code
$$
C^{\bot_{k}}=\big\{{\bf x}\in\mathbb{F}_{q} ^n\,\big|\,
 [{\bf c},{\bf x}]_{k}=0,\, \forall~{\bf c}\in C\big\}
$$
is called the $k$-Galois dual code of $C$.
Note that $C^{\bot_{k}}$ is linear whenever $C$ is linear or not.
Then $C^{\bot_{0}}$ (simply, $C^{\perp}$) is just the Euclidean dual code of $C$,
and $C^{\bot_{\frac{e}{2}}}$ (simply, $C^{\perp_{H}}$)
is just the Hermitian dual code of $C$.
If $C \subset C^{\bot_{k}}$, then $C$ is said to be $k$-Galois self-orthogonal.
Moreover, $C$ is said to be $k$-Galois self-dual if $C= C^{\bot_{k}}$.

From the fact that the $k$-Galois form is non-degenerate (\cite[Remark 4.2]{FanY}),
it follows immediately that
$\dim_{\mathbb{F}_{q}}C+\dim_{\mathbb{F}_{q}}C^{\perp_{k}}=n$.

\begin{define}\rm
A  linear code $C$ over $\mathbb{F}_{q}$ is called a
{\em linear complementary $k$-Galois dual code}
(abbreviated to {\em $k$-Galois LCD code})
if $C^{\perp_{k}}\cap C=\{\mathbf{0}\}$.
\end{define}

For an $s\times s$ matrix $A =(a_{ij})_{s\times s}$ over $\mathbb{F}_{q}$,
let $A^{(p^{e-k})} := (a_{ij}^{p^{e-k}})_{s\times s}$. Denote by $A^{\ddagger}$ the transpose matrix of $A^{(p^{e-k})}$.
For a vector $\mathbf{a}=(a_1,a_2,\ldots,a_n)\in\mathbb{F}_{q} ^n$,
we have
$$\mathbf{a}^{p^{e-k}}=(a_1^{p^{e-k}},a_2^{p^{e-k}},\ldots,a_n^{p^{e-k}}).$$

For a linear code $C$ of $\mathbb{F} _{q}^n$,
we define $C^{p^{e-k}}$ to be the set
$\{\mathbf{a}^{p^{e-k}}\mid~\mathbf{a}\in C\}$ which is also a linear code.
It is easy to see that the $k$-Galois dual $C^{\perp_{k}}$ of $C$ is equal to the
Euclidean dual $(C^{p^{e-k}})^{\perp}$ of the linear code $C^{p^{e-k}}$.
We write these easy facts in the following lemma.

\begin{lemma} \label {le:2.2}
Let $C$ be an $[n,l,d]_q$ linear code over $\mathbb{F}_{q}$ with a generator matrix $G$.
Then $C^{p^{e-k}}$ is an $[n,l,d]_q$ linear code  over $\mathbb{F}_{q}$
with a generator matrix $G^{(p^{e-k})}$;
and $(C^{p^{e-k}})^{\perp}=C^{\perp_{k}}$, hence,
$C$ is $k$-Galois LCD if and only if
$C \cap (C^{p^{e-k}})^{\perp}=\{\mathbf{0}\}$.
\end{lemma}

The following fact is well known, e.g., see \cite{Liu2}.

\begin{prop} \label {pro:2.3}Let $C$ be a linear code over $\mathbb{F}_{q}$.
The following two statements are equivalent:

$\mathrm{(1)}$ $C$ is MDS;

$\mathrm{(2)}$ $C^{\perp_{k}}$ is MDS.
\end{prop}

The following theorem shows a criteria of $k$-Galois LCD codes which can be found in \cite{Liu2}.

\begin{theorem} \label {th:2.4}
Let C be an $[n,l,d]_q$ linear code over $\mathbb{F} _{q}$ with generator matrix $G$.
Then $C$ is  $k$-Galois LCD if and only if $GG^{\ddagger}$ is nonsingular.
\end{theorem}

\subsection{Entanglement-assisted quantum codes}
In this subsection, we first recall some basic concepts and results about entanglement-assisted quantum codes (see\cite{Fuji},\cite{Bowen01},\cite{Brun01},\cite{Lai01},\cite{Wilde}).

An $[[n,k,d;c]]_q$ entanglement-assisted quantum error-correcting code (EAQECC) over $\mathbb{F}_{q}$ encodes $k$ logical qubits into $n$ physical qubits with the help of $c$ copies of maximally entangled states ($c$ ebits). The performance of an EAQECC is measured by its rate $\frac{k}{n}$ and net rate $\frac{k-c}{n}$.
If $c=0$, then the EAQECC is a standard stabilizer code.  EAQECCs  can be regarded as generalized quantum codes.

It has been prove that EAQECCs have some advantages over standard stabilizer codes. In \cite{Wilde}, Wilde and Brun proved that EAQECCs can be constructed using classical linear codes as follows.

\begin{prop} \label{prop:2.1} (\cite{Wilde})
Let $H_{1}$  and $H_2$ be parity check matrices of two linear codes $[n,k_1,d_1]_q$ and $[n,k_2,d_2]_q$,  respectively. Then an $[[n,k_1+k_2-n+c,\mathrm{min}\{d_1,d_2\};c]]_q$ EAQECC can be obtained, where $c=\mathrm{rank}(H_1H_2^{T})$ is the required number of maximally entangled states.
\end{prop}

To see that an EAQECC is good in terms of its parameters, we have to introduce the entanglement-assisted quantum Singleton bound.
\begin{theorem} \label{theorem:2.1}(\cite{Brun01})
Let $Q$ be an EAQECC with parameters $[[n,k,d;c]]_{q}$. Then $2(d-1)\leq n-k+c$, where $0\leq c\leq n-1$.
\end{theorem}

If  an  EAQECC  $Q$ with parameters $[[n,k,d;c]]_{q}$ attains the entanglement-assisted quantum Singleton bound $2(d-1)= n-k+c$, then it is called a maximum-distance-separable  EAQECC (MDS EAQECC).

\begin{define}\label {de:2.2}
Let $Q$ be an EAQECC with parameters $[[n,k,d;c]]_{q}$. If  $c=n-k$, it is called a maximal-entanglement EAQECC.
\end{define}

\section{Construction of EAQECC from $k$-Galois LCD codes}
In this  section, we give four classes of Entanglement-assisted quantum codes from $k$-Galois LCD codes. All of resulting codes are MDS.
\begin{lemma}\label{le:3.1} Let $C$ be a linear $[n,l,d]_q$ code with parity check matrix $H$. Then
$$rank(HH^{\ddagger})=n-l-dim(C\cap C^{\perp_k}).$$
Furthermore, if $C$ is a  $k$-Galois LCD code, then $C^{\perp_k}$ is also a $k$-Galois LCD code.
\end{lemma}
\pf Let $t=dim(C\cap C^{\perp_k})$ and $A=\{h_1,\ldots,h_t\}$ be a basis of $C\cap C^{\perp_k}$. Extend $A$ to be a basis $\{h_1,\ldots,h_t,h_{t+1},\ldots,h_{n-l}\}$ of $C^{\perp_k}$.  Let
$$M=\begin{pmatrix}h_{1}\\ h_{2}  \\ \vdots \\h_{n-l}\end{pmatrix}.$$
It is easy to see that $M^{(p^k)}$ is a parity check matrix of $C$. After a suitable sequence of elementary  row operations, we have that $H=BM^{(p^k)}$ for some invertible $(n-l)\times(n-l)$ matrix $B$ over $\mathbb{F}_q$, and we have
$$HH^{\ddagger}=BM^{(p^k)}M^{T}B^{\ddagger}.$$
By $B$ and $B^{\ddagger}$ are invertible, we have
$$rank(HH^{\ddagger})=rank(M^{(p^k)}M^{T})=n-l-t=n-l-dim(C\cap C^{\perp_k}).$$

Since  $C$ is a  $k$-Galois LCD code, $\mathrm{dim}(C\cap C^{\perp_k})=0$, which implies that $rank(HH^{\ddagger})=n-l$. Therefore,
$$rank(H^{(p^{e-k})}(H^{(p^{e-k})})^{\ddagger})=rank(HH^{\ddagger})^{(p^{e-k})}=n-l.$$
 In light of Theorem  \ref{th:2.4},  $C^{\perp_k}$ is also a  $k$-Galois LCD code.
\qed

\begin{remark} Generally,  $(C^{\perp_k})^{\perp_k}\neq C$ for some $0\leq k <e$.
\end{remark}

The following Proposition will play an important role in constructing MDS EAQECCs.
\begin{prop}\label{pr:3.2} Let $C$ be an $[n,l,d]_q$  $k$-Galois LCD  code with parity check matrix $H$. Then

$(1)$ there exists an $[[n,l,d;n-l]]_q$  EAQECC.

$(2)$ if $C$ is a $k$-Galois LCD  MDS code ,  then there exists an $[[n,l,d;n-l]]_q$ MDS EAQECC.
\end{prop}
\pf  $(1)$ By Lemma \ref{le:3.1} and Proposition \ref {pro:2.3} , we know that $C^{\perp_k}$ is an $[n,n-l,l+1]_q$  $k$-Galois LCD  code.

It is easy to prove that $(C^{\perp})^{p^{e-k}}=(C^{p^{e-k}})^{\perp}$.

Take $H_1=H$ and $H_2=H^{(p^{e-k})}$,  then $H_1$ and $H_2$ are parity check matrices of the linear codes $C$ and $C^{p^{e-k}}$, respectively.

By Proposition \ref {prop:2.1}, we obtain an $[[n,l,d;n-l]]_q$ EAQECC.

$(2)$ According  to  the assumption and Theorem \ref{theorem:2.1}, it is easy to prove that there exists an $[[n,l,d;n-l]]_q$ MDS EAQECC.
\qed

\subsection{The first classes of MDS EAQECCs}
The following theorem had be showed by \cite{Liu2}.
\begin{theorem}\label{th:3.3}
Let $\gcd(p^{e-k}+1,p^e-1)=s$ and $p^e-1=st$.
If $0\leq l<\min\{t,n\} $ and $ n\leq q+1$,
then there exist $k$-Galois LCD MDS $[n,l,n-l+1]_q$ codes
and $k$-Galois LCD MDS $[n,n-l,l+1]_q$ codes.
\end{theorem}

Combing  Proposition \ref{pr:3.2} and Theorem \ref{th:3.3},  we can obtain the following MDS EAQECCs with length $n\leq q+1$.
\begin{theorem} \label{th:B}
Let $\gcd(p^{e-k}+1,p^e-1)=s$ and $p^e-1=st$.
If $0\leq l<\min\{t,n\} $ and $ n\leq q+1$,
then there exist  $[[n,l,n-l+1;n-l]]_q$ MDS EAQECCs
and  $[[n,n-l,l+1;l]]_q$ MDS EAQECCs.
\end{theorem}
\begin{remark} Since $c=n-l$ or  $c=l$,   it follows that the constructed MDS EAQECCs by Theorem \ref{th:B} are also maximal-entanglement EAQECCs. On other the hand, $k$-Galois LCD codes include the usual LCD codes and Hermitian LCD codes as two special cases.
The more general setting allows us to find more MDS EAQECCs.
\end{remark}

\begin{example} Let $q=5^2,n=q+1=26$. Take $k=1$, then $\gcd(p^{e-k}+1,p^e-1)=\gcd(6,24)=6$. Applying  Theorem \ref{th:B}, we obtain eight MDS EAQECCs whose parameters are $[[26,1,26;25]]_{25},[[26,2,25;24]]_{25},[[26,3,24;23]]_{25},[[26,4,23;22]]_{25},[[26,25,2;1]]_{25},[[26,24,3;2]]_{25}$,\\$[[26,23,4;3]]_{25},[[26,22,5;4]]_{25}$.
\end{example}

\subsection{The second classes of MDS EAQECCs}
In this subsection, we first recall some basic results of generalized Reed-Solomon codes (see \cite{JLLX10} and \cite{JX14}). Let $a_1,\ldots,a_n$ be $n$ distinct elements of $\mathbb{F}_{q}$, and let  $v_1,\ldots,v_n$ be $n$ nonzero elements of $\mathbb{F}_{q}$. For $l$ between $1$ and $n$, the generalized Reed-Solomon code $GRS_l(\mathbf{a},\mathbf{v})$ is defined by
$$GRS_l(\mathbf{a},\mathbf{v})=\{(v_1f(a_1),\ldots,v_nf(a_n))|~f(x)\in\mathbb{F}_{q}[x],deg(f(x))\leq l-1\},$$
where $\mathbf{a}$, and $\mathbf{v}$ denote the vectors $(a_1,\ldots,a_n)$ and $(v_1,\ldots,v_n)$, respectively.
The generalized Reed-Solomon code $GRS_l(\mathbf{a},\mathbf{v})$ is an $[n, l]$ MDS code and it has a generator
matrix
$$ G_{(\mathbf{a},\mathbf{v})}=\begin{pmatrix}v_1&v_2&\cdots&v_n\\v_1a_1&v_2a_2&\cdots&v_na_n\\v_1a_1^2&v_2a_2^2&\cdots&v_na_n^2\\ \vdots&\vdots&\ddots&\vdots\\v_1a_1^{l-1}&v_2a_2^{l-1}&\cdots&v_na_n^{l-1}\end{pmatrix}.$$

Define an extended generalized Reed-Solomon code:
$$GRS_l(\mathbf{a},\mathbf{v},\infty)=\{(v_1f(a_1),\ldots,v_nf(a_n),f_{l-1})|~f(x)\in\mathbb{F}_{q}[x],deg(f(x))\leq l-1\},$$
where $f_{l-1}$ is the coefficient of $x^{l-1}$ in $f (x)$. The extended generalized Reed-Solomon code $GRS_l(\mathbf{a},\mathbf{v},\infty)$ is an $[n + 1, l,n-l+2]$ MDS code and it has a generator matrix
$$ G_{(\mathbf{a},\mathbf{v},\infty)}=\begin{pmatrix}v_1&v_2&\cdots&v_n&0\\v_1a_1&v_2a_2&\cdots&v_na_n&0\\v_1a_1^2&v_2a_2^2&\cdots&v_na_n^2&0\\ \vdots&\vdots&\ddots&\vdots&\vdots\\v_1a_1^{l-1}&v_2a_2^{l-1}&\cdots&v_na_n^{l-1}&1\end{pmatrix}.$$

Next, we  need a helpful lemma with respect to  $k$-Galois LCD MDS codes.

\begin{lemma} \label{le:3.4} Let $q=p^e$ be odd, $l$ be a positive integer with $l\mid(p^e- 1), l\nmid p^k + 1, 2(p^k+1)\mid(p^e- 1)$, and $l\leq\frac{p^e-1}{2}$. Then there are $[2l,l,l+1]_q,[2l + 1,l,l+2]_q$ and $[2l+ 2,l,l+3]_q$  $k$-Galois LCD MDS code over $\mathbb{F}_q$.
\end{lemma}
\pf Let $\alpha$ be   a primitive element of $\mathbb{F}_q$. Then $\alpha^{\frac{p^e-1}{l}}$ generates a subgroup with order $l$ of $\mathbb{F}_q^*$. Suppose that the subgroup $\langle\alpha^{\frac{p^e-1}{l}}\rangle=\langle\beta_0,\beta_1,\ldots\beta_{l-1}\rangle$, where $\beta_j\in \mathbb{F}_q^*$ and $\beta_0 = 1$.
For any integer $s$ and any $\delta \in \mathbb{F}_q^*$, one has
$$(\delta\beta_0)^s+(\delta\beta_1)^s+\ldots+(\delta\beta_{l-1})^s=\begin{cases}
         \delta^sl, &~s\equiv0~\mathrm{mod}~l;\\
         0, &\mathrm{otherwise}.
         \end{cases} $$

Then

$$(\delta\beta_0)^i({\delta\beta_0})^{p^kj}+(\delta\beta_1)^i({\delta\beta_1})^{p^kj}+\ldots+(\delta\beta_{l-1})^i({\delta\beta_{l-1}})^{p^kj}=\begin{cases}
         \delta^{i+p^kj}l, &~i+p^kj\equiv0~\mathrm{mod}~l;\\
         0, &\mathrm{otherwise}.
         \end{cases} $$

Since  $2(p^k+1)\mid(p^e- 1)$, we can take that $\gamma=\frac{1}{\alpha^{\frac{p^e-1}{2(p^k+1)}}}$.

It is  easy to prove that $\beta_0,\ldots,\beta_{l-1},\alpha\beta_0,\ldots,\alpha\beta_{l-1}\in \mathbb{F}_q$ are distinct elements

Cases 1: Take $\mathbf{a} = (\beta_0,\ldots,\beta_{l-1},\alpha\beta_0,\ldots,\alpha\beta_{l-1})$  and $\mathbf{v}=(1,\ldots,1,\gamma,\ldots\gamma)$.  Then the generalized Reed-Solomon code $GRS_l(\mathbf{a},\mathbf{v})$  generated
by matrix $ G_{(\mathbf{a},\mathbf{v})}$ is an MDS code.

Let $a(i,j)$ be the entry in the $i$-th row and $j$-th column of the matrix
$G_{(\mathbf{a},\mathbf{v})}G_{(\mathbf{a},\mathbf{v})}^{\ddagger}$, where $i,j=0,1,\ldots,l-1$. Then
$$a(i,j)=\begin{cases}
         (1-\frac{\alpha^{i+p^kj}}{\alpha^{1+p^k}})l, &~i+p^kj\equiv0~\mathrm{mod}~l;\\
         0, &\mathrm{otherwise}.
         \end{cases} $$
When $i+p^kj\equiv0~\mathrm{mod}~l$, we assume that $a(i,j) = 0$. Then $\alpha^{i+p^kj} = \alpha^{1+p^k}$ and $i + p^kj \equiv 1 +p^k~\mathrm{mod }~(p^e-1)$. From $l|(p^e-1)$ and $l|(i + p^kj)$, we have $l|(1+p^k)$, which implies a contradiction with $l\nmid (1+p^k)$. Thus, when $i+p^kj\equiv0~\mathrm{mod}~l$, we have
$a(i,j)\neq 0$. Note that $l|(p^e- 1)$. Every row of $G_{(\mathbf{a},\mathbf{v})}G_{(\mathbf{a},\mathbf{v})}^{\ddagger}$ has just only a nonzero element. That holds for each column.
Hence, the matrix $G_{(\mathbf{a},\mathbf{v})}G_{(\mathbf{a},\mathbf{v})}^{\ddagger}$ is nonsingular. From Theorem \ref{th:2.4}, $GRS_l(\mathbf{a},\mathbf{v})$ is an $[2l,l,l+1]_q$ $k$-Galois LCD MDS code.

Cases 2: Similar to the cases 1, take $\mathbf{a} = (\beta_0,\ldots,\beta_{l-1},\alpha\beta_0,\ldots,\alpha\beta_{l-1})$  and $\mathbf{v}=(1,\ldots,1,\gamma,\ldots\gamma)$.  Then the extended generalized Reed-Solomon code $GRS_l(\mathbf{a},\mathbf{v},\infty)$  generated
by matrix $ G_{(\mathbf{a},\mathbf{v},\infty)}$ is an MDS code. Obviously, $GRS_l(\mathbf{a},\mathbf{v},\infty)$ is an $[2l+1,l,l+2]_q$ $k$-Galois LCD MDS code.

Cases 3: Since $q > 2$, there exists an $\eta \in \mathbb{F}_q^*$ such that $\eta\eta^{p^{e-k}}\neq(1-\frac{1}{\alpha^{1+p^k}})l$.  Let $C_3 $ be a linear code generated by the following matrix
$$G_3=(G_{(\mathbf{a},\mathbf{v})}\mid\eta e_0\mid e_l).$$
where $e_0 = [1,0,\ldots ,0]^T$ and $e_l = [0,0,\ldots ,1]^T$. From the  discussion in cases 1, we have that $C_3$ is an
$[2l + 2,l,l+3]_q$ $k$-Galois LCD MDS code.

This proves the expected results.
\qed

Combing  Proposition \ref{pr:3.2} and Lemma \ref{le:3.4},  we can obtain the following MDS EAQECCs.
\begin{theorem} \label{th:C}
Let $q=p^e$ be odd, $l$ be a positive integer with $l\mid(p^e- 1), l\nmid p^k + 1, 2(p^k+1)\mid(p^e- 1)$, and $l\leq\frac{p^e-1}{2}$.
Then there exist  $[[2l,l,l+1;l]]_q$, $[[2l+1,l,l+2;l+1]]_q$,
and  $[[2l+2,l,l+3;l+2]]_q$ MDS EAQECCs.
\end{theorem}
\begin{remark} It is easy to see that the constructed MDS EAQECCs by Theorem  \ref{th:C} are also maximal-entanglement EAQECCs.
\end{remark}

\begin{example} Let $p=7,e=12$, i.e., $q=7^{12}$. Then
$$7^{12}-1=(7^6-1)(7^6+1)$$
$$~~~~~~~~~~~~~~~~=(7^3-1)7^3+1)(7^6+1)$$
$$~~~~~~~~~~~~~~~~~~~~~~~~~~~~~~~~~~~~~~~=(7+1)(7^2+7+1)(7^3+1)(7^3-1)(7^6+1)$$
$$~~~~~~~~~~~~~~~~~~~~~~~=(7^2-1)(7^2+1)(7^8+7^4+1).$$
We consider the following several cases.

Cases 1. Take $k=1, l=2293469100$. Then  $l\mid(p^e- 1), l\nmid p^k + 1, 2(p^k+1)\mid(p^e- 1)$. Applying  Theorem \ref{th:C}, we obtain three MDS EAQECCs whose parameters are $$[[4586938200,2293469100,2293469101;2293469100]]_{7^{12}},$$$$[[4586938201,2293469100,2293469102;2293469101]]_{7^{12}},$$$$[[4586938202,2293469100,2293469103;2293469102]]_{7^{12}}.$$

Cases 2. Take $k=2, l=48$. Then  $l\mid(p^e- 1), l\nmid p^k + 1, 2(p^k+1)\mid(p^e- 1)$. Applying  Theorem \ref{th:C}, we obtain three MDS EAQECCs whose parameters are $~[[96,48,49;48]]_{7^{12}},$
$[[97,48,50;49]]_{7^{12}}, [[98,48,51;50]]_{7^{12}}.$

Cases 3. Take $k=3, l=342$. Then  $l\mid(p^e- 1), l\nmid p^k + 1, 2(p^k+1)\mid(p^e- 1)$. Applying  Theorem \ref{th:C}, we obtain three MDS EAQECCs whose parameters are $[[684,342,343;342]]_{7^{12}},$
$[[685,342,344;343]]_{7^{12}}, [[686,342,345;344]]_{7^{12}}.$

Cases 4. Take $k=6, l=117648$. Then  $l\mid(p^e- 1), l\nmid p^k + 1, 2(p^k+1)\mid(p^e- 1)$. Applying  Theorem \ref{th:C}, we obtain three MDS EAQECCs whose parameters are $[[235296,117648,117649;$
$117648]]_{7^{12}},[[235297,117648,117650;117649]]_{7^{12}}, [[235298,117648,117651;117650]]_{7^{12}}.$
\end{example}

\begin{remark}  From above examples, if we take different values of $k$ and $l$, then three MDS EAQECCs
 can be constructed by Theorem \ref{th:C}. Therefore, for a fixed $q$,  many MDS EAQECCs can be constructed from  Theorem \ref{th:C}.
\end{remark}

\subsection{The third classes of MDS EAQECCs}
We  recall a fact from  constacyclic codes which can be found in \cite{FanY}.
\begin{lemma} \label{le:3.2}If $C$ is a $\lambda$-constacyclic code of length $n$ over $\mathbb{F}_{q}$, then the $k$-Galois dual code
$C^{\perp_{k}}$ is a $\lambda^{-p^{e-k}}$-constacyclic code
of length $n$ over $\mathbb{F}_{q}$ .
\end{lemma}
A sufficient condition for $\lambda$-constacyclic codes to be $k$-Galois LCD codes can be found in \cite{Liu2}.

\begin{cor} \label{cor:3.3} Let $\lambda \in \mathbb{F}_{q}^{*}$ be a primitive
r-th root of unity. If $\lambda^{1+p^{e-k}}\neq1$, i.e., $r\mid (p^e-1)$, $r\nmid(1+p^{e-k})$, then any $\lambda$-constacyclic $C$ of length $n$ over $\mathbb{F}_{q}$ is a $k$-Galois LCD code.
\end{cor}

\begin{theorem} \label{th:A} Let $q=p^e$ where $p$ is an odd prime. Let $\lambda \in \mathbb{F}_{q}^{*}$ be a primitive
r-th root of unity, where $r|(q-1)$ and $r \nmid(1+p^{e-k})$. Set $n=\frac{q-1}{r}$. Then

$\mathrm{(1)}$ any $\lambda$-constacyclic code is $k$-Galios LCD code.

$\mathrm{(2)}$ for any $2\leq d \leq \frac{q-1}{r}$, there  exists an $[[n,n+1-d,d;d-1]]_q$ MDS EAQECC.
\end{theorem}
\pf $\mathrm{(1)}$ By the assumption and Corollary \ref{cor:3.3}, the statement is obvious.

$\mathrm{(2)}$ Since $n=\frac{q-1}{r}$, we have $q \equiv 1~\mathrm{mod}~rn$. Thus, each $q$-cyclotomi coset modulo $rn$ only has one element. Set $C_{1+rj}=\{1+rj\}$ for $0\leq j\leq n-1$. Therefore, $1+r\mathbb{Z}_{rn}=\cup_{j=0}^{n-1}C_{1+rj}$.

Let us consider the $\lambda$-constacyclic $k$-Galois LCD code $C$ with defining set
$$P = \cup_{j=0}^{d-2} C_{1+rj}.$$
where $2\leq d\leq n$. Obviously, the defining set $P$ of $C$  consists of $d-1$ consecutive integers $\{1,1+r,1+2r,\ldots,1+(d-2)r\}$. Thus  $C$ has dimension $l =n+1-d$. From the BCH bound  for constacyclic codes, the minimum distance of $C$ is at least $d$.
Applying the classical code Singleton bound to $C$ yields  a
$k$-Galois LCD  MDS code with parameters $[n,n+1-d,d]_q$.

In light of Proposition \ref{pr:3.2},  there  exists an $[[n,n+1-d,d;d-1]]_q$ MDS EAQECC  for $2\leq d \leq n$.
\qed

\begin{example} Let $q=7^3=343,k=1$.

$(1)$ Take $r=3$. Then $n=114$. Obviously, $3\mid(7^3-1)$ and $3 \nmid(1+7^{2})$.  Applying  Theorem \ref{th:A}, we obtain 113 MDS EAQECCs whose parameters are $[[114,115-d,d;d-1]]_{343}$ for $2\leq d \leq 114$.

$(2)$ Take $r=6$. Then $n=57$. Obviously, $6\mid(7^3-1)$ and $6 \nmid(1+7^{2})$.   Applying  Theorem \ref{th:A}, we obtain 56 MDS EAQECCs whose parameters are $[[57,58-d,d;d-1]]_{343}$ for $2\leq d \leq 57$.
\end{example}
\begin{example} Let $q=13^3=2197,k=1$. Then $n=732$.  Obviously, $3\mid(13^3-1)$ and $3 \nmid(1+13^{2})$.  Applying  Theorem \ref{th:A}, we obtain 731 MDS EAQECCs whose parameters are $[[732,733-d,d;d-1]]_{2197}$ for $2\leq d\leq 732$.
\end{example}

\subsection{The fourth classes of MDS EAQECCs}
Note that, if $\lambda^{1+p^{e-k}}=1$, then the $k$-Galois dual
$C^{\perp_k}$ of any $\lambda$-constacyclic code $C$ is still $\lambda$-constacyclic.
The following  useful criterion to test whether
a $\lambda$-constacyclic code is a $k$-Galois LCD code can be found in  \cite{Liu2}.

\begin{theorem} \label{th:3.4}
Assume that $\lambda^{1+p^{e-k}}=1$.
Let $P\subseteq 1+r\mathbb{Z}_{rn}$ be a union of some $q$-cosets,
 and
$C_P\subseteq R_{n,\lambda}$ be the $\lambda$-constacyclic code
with defining set $P$. Then $C_P$ is a $k$-Galois LCD $\lambda$-constacyclic  code if and only if $-p^{k}P=P$.
\end{theorem}

According to above theorem, we gave two subclasses of MDS EAQECCs with small length by mean of the following lemma in \cite{Liu2}.

\begin{lemma}\label{le:3.5} Let $q=p^e$ with $e\geq2$, let $k=1$.

{\rm(1)}~ If $p\equiv1\pmod{4}$, then there exists a
$1$-Galois LCD MDS negacyclic code over $\mathbb{F}_{q}$
with parameters $[(p-1)/2,(p-1)/2-2l,2l+1]_q$, where $1\leq l\leq(p-5)/4$.

{\rm(2)}~ If $p\equiv3\pmod{4}$, then there exists a
$1$-Galois LCD MDS negacyclic code over $\mathbb{F}_{q}$
with parameters $[(p-1)/2,(p-1)/2-2l+1,2l]_q$, where $1\leq l\leq(p-3)/4$.
\end{lemma}

Combing  Proposition \ref{pr:3.2} and Lemma \ref{le:3.5},  we can obtain the following MDS EAQECCs with length $n=p$.

\begin{cor}\label{cor:3.5} Let $q=p^e$ with $e\geq2$.

{\rm(1)}~ If $p\equiv1\pmod{4}$, then there exists a
 MDS EAQECC over $\mathbb{F}_{q}$ with parameters $[[(p-1)/2,(p-1)/2-2l,2l+1;2l]]_q$, where $1\leq l\leq(p-5)/4$.

{\rm(2)}~ If $p\equiv3\pmod{4}$, then there exists a
MDS EAQECC over $\mathbb{F}_{q}$ with parameters $[[(p-1)/2,(p-1)/2-2l+1,2l;2l-1]]_q$, where $1\leq l\leq(p-3)/4$.
\end{cor}

\section{Conclusion}
We give four classes of  MDS  EAQECCs from $k$-Galois LCD MDS codes for some certain code lengths.  From our results, it can be seen a class of MDS  EAQECCs  can be obtained if we fined a class of $k$-Galois MDS LCD codes.  We believe that more MDS EAQECCs can be obtained from $k$-Galois MDS LCD codes.

\textbf{Acknowledgements}

The authors would like to sincerely thank the editor and the referees
for very meticulous readings of this paper,
and for valuable suggestions which help us to create an improved version.
X. Liu was supported by Research Funds of Hubei Province
(Grant No. D20144401 and Q20174503), and Research Project of
Hubei Polytechnic University (Grant No. numbers 17xjz03A ).
H. Liu was supported by China Scholarship Council (Grant No. 201606770024),
and the Educational Commission of Hubei Province (Grant No. B2015096).

\end{document}